\documentstyle[11pt,newpasp,twoside,psfig]{article}
\begin{document}

\title{AGN fueling: the observational point of view}
\author{Fran\c{c}oise Combes}
\affil{Observatoire de Paris, 61 Av. de l'Observatoire,
F-75 014, Paris, France}

\setcounter{page}{111}
\index{Combes, F.}

\begin{abstract}
Observations at multiple wavelengths are reviewed to search 
for evidence for fueling mechanisms in galaxies,
both for nuclear starbursts and AGN activity. Although it is
 undisputed that dynamical perturbations such as
bars or tidal interactions accumulate gas in the central regions
 and sometimes trigger nuclear starbursts, the
evidence remains scarce that these are necessary to fuel AGNs. 
Interpretations in terms of time-scales,
feed-back, and black hole evolution are discussed. It is suggested
that the AGN phase corresponds to the early-type phases of galaxies.
\end{abstract}

\section{Fueling mechanisms}

The essential issue to fuel gas into nuclei is to find
efficient processes to
transfer angular momentum. Since viscous torques are
not efficient except very close to the
accretion disk itself, the principal actors here are
gravity torques due to non-axisymmetric structures:\\
\indent 1. Bars, spirals, m=1 waves\\
\indent 2. Galaxy interactions and mergers\\
\noindent These mechanisms are indeed observed when the AGN activity is very strong
and accretion rate exceptional ($>$ 10 M$_\odot$/yr).
For milder activities, it is difficult to see clear correlation
with bars or companions. Nuclear
starbursts and AGN have similar fueling requirements: they compete
for gas supply and occur often simultaneously.
The nuclear activity is however much more efficient to
radiate energy and the 
fraction of energy contributed by the AGN (with 
respect to starbursts) increases with total luminosity.

\section{Correlation with bars}

\subsection{Primary bars}

Many studies have addressed the role of bars in nuclear activity
(Simkin et al 1980, Dahari 1984, Moles et al 1995).
Weak correlations are found in general, and it is often
difficult to discriminate the role of biases due to
morphological types, environment, etc..
Bars are more easily detected in the old stellar component,
and in near infrared surveys, free of dust extinction: 
many more barred galaxies are discovered then, but Seyferts
are not privileged (McLeod \& Rieke 1995, Mulchaey \& Regan 1997).
The bottom line of most surveys is that there is
no difference of bar frequency 
between AGN and non-active galaxies (see fig. 1).

\begin{figure}[t]
\centerline{
\psfig{figure=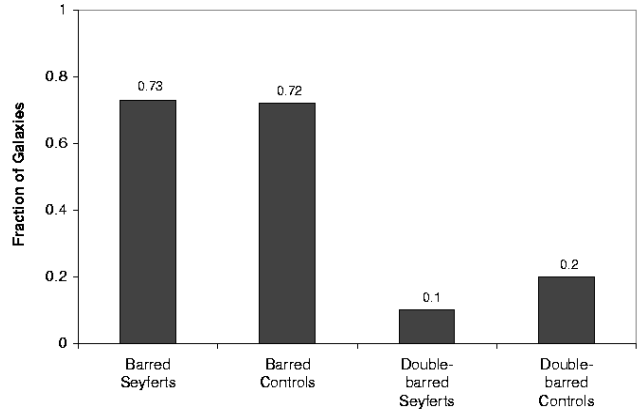,width=9cm,angle=0}
}
\caption{Fraction of barred and double-barred systems in the Seyfert and
control galaxy samples. It is not possible to distinguish
the two samples with the percentage of barred galaxies.
Slightly more normal galaxies have double bars (from Mulchaey \& Regan 1997) }
\label{fig1}
\end{figure}

A recent study comparing the CfA sample of Seyferts to a control sample
finds slightly more bars in Seyferts (Peletier et al 1999, Knapen et al 2000):
79\% + 7.5\% of Seyferts are barred, while
59\% + 9\%  control galaxies are barred.
Paradoxically, there is also a
lower fraction of strong bars in  Seyferts with respect to the
control sample (Shlosman et al 2000). This could be 
the consequence of bar destruction by central masses.
As expected from gravity torques and consequent radial gas inflow,
barred galaxies do possess larger molecular
gas concentrations, revealed by CO 
observations (Sakamoto et al 1999). This gas inflow
produces more nuclear starbursts, which 
correspond to conspicuous circumnuclear HII hot spots (Sersic 1963).

\subsection{Decoupled stellar nuclear disks}

However, the gas infall may stop at the inner Lindblad resonance
of the large bar, and thus be prevented to fuel the nucleus. Are
then nuclear bars the final processes to fuel the central regions?
With high spatial resolution, it is now possible to investigate
morphological features inside the central kiloparsec in galaxies.
The first studies do not find more nuclear bars in Seyferts
(e.g. Regan \& Mulchaey 1999); the
frequency of nuclear bars is 20-30\%, the same in the control sample
and for the AGN samples. Laine et al (2002) find a larger fraction 
of nested bars in Seyfert2 with respect to Seyfert1.

A striking feature is the high frequency
of nuclear spirals discovered (Martini \& Pogge 1999, 
Pogge \& Martini 2002).  The frequent presence of 
nuclear disks (dust, gas and young stars) point towards the action
of bars and interactions, to produce the gravity torques and
matter inflow. Emsellem et al (2001) discovered 
cold stellar nuclear disks in double-barred Seyfert galaxies,
related to this inflow process. The observed stellar velocity
dispersion reveals a characteristic drop in the center,
interpreted in terms of recent nuclear star formation, as supported
by numerical simulations (Wozniak  et al 2002).

\subsection{Why is there no correlation?}

Many arguments could be advanced
to explain the absence of strong correlation between
bars and AGNs (e.g. Combes 2001). First, even if the bar produces gas inflow,
there must be a sufficiently massive black hole at the nucleus
for an AGN to appear, and this is not everywhere, in particular
in view of the mass correlation between black holes and bulges:
the condition is not fulfilled in late-type galaxies. 
Second, there could be a stellar bar, but with insufficient gas content
to produce a strong enough fueling. Third, the time-scales
for bar formation (and destruction), the gas fueling, and 
the subsequent nuclear activity could be different, and the
phenomena could occur at slightly different epochs.

An interesting feature relating bars and nuclear activity is that
Seyferts have more outer rings (by a factor 3-4) than 
non-active galaxies; these outer rings are thought to be the vestige of
previous bars (Hunt \& Malkan 1999). In their extended 12 microns sample 
of 891 galaxies, Hunt \& Malkan (1999) find that  
30\% are AGN, 25\% are interacting (perturbed).
While HII/Starburst galaxies have more bars,
AGN (LINERs and Seyferts) have no more bars but rings.
LINERs have more inner rings, while Seyferts have
3-4 times more outer rings (cf fig. 2).

\begin{figure}[t]
\centerline{
\psfig{figure=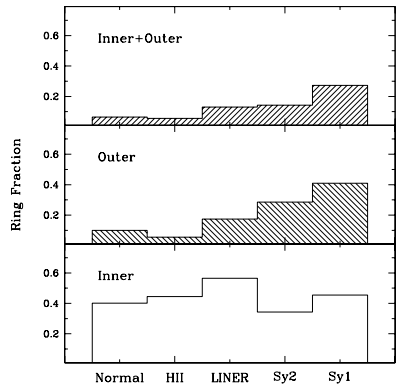,width=6cm,angle=0}
}
\caption{Fraction of inner (usually surrounding  bars) and outer
ring galaxies as a function of activity class.
Only objects with ($v\,<$\,5000\,km\,s$^{-1}$) are shown here
(from Hunt \& Malkan, 1999)}
\label{fig2}
\end{figure}

\section{Statistics about AGN}

AGN tend to lie in early-type galaxies (Terlevich et al 1987, Moles et al 1995).
In an optical spectroscopic survey of 486 nearby galaxies, Ho et al (1997) detected 
420 emission-lines nuclei (86\% detection rate). Half of these objects can
be classified as HII or star-forming nuclei, and half as some kind of AGN: Seyfert,
LINERs and transition objects LINER/HII. A signature of Broad Line Region is
found in 20\% of the AGN, while Seyfert nuclei reside in about 10\% of
all galaxies. AGNs are found predominantly in luminous, early-type galaxies,
while HII nuclei are in less luminous late-type objects (see fig. 3).
The relation between bulge mass and black hole
mass might be the reason why AGN 
are favored in early-types. Also the concentration
of mass in the bulge favors the 
presence of an ILR, then nuclear bar and leads to easiest fueling.

\begin{figure}[t]
\centerline{
\psfig{figure=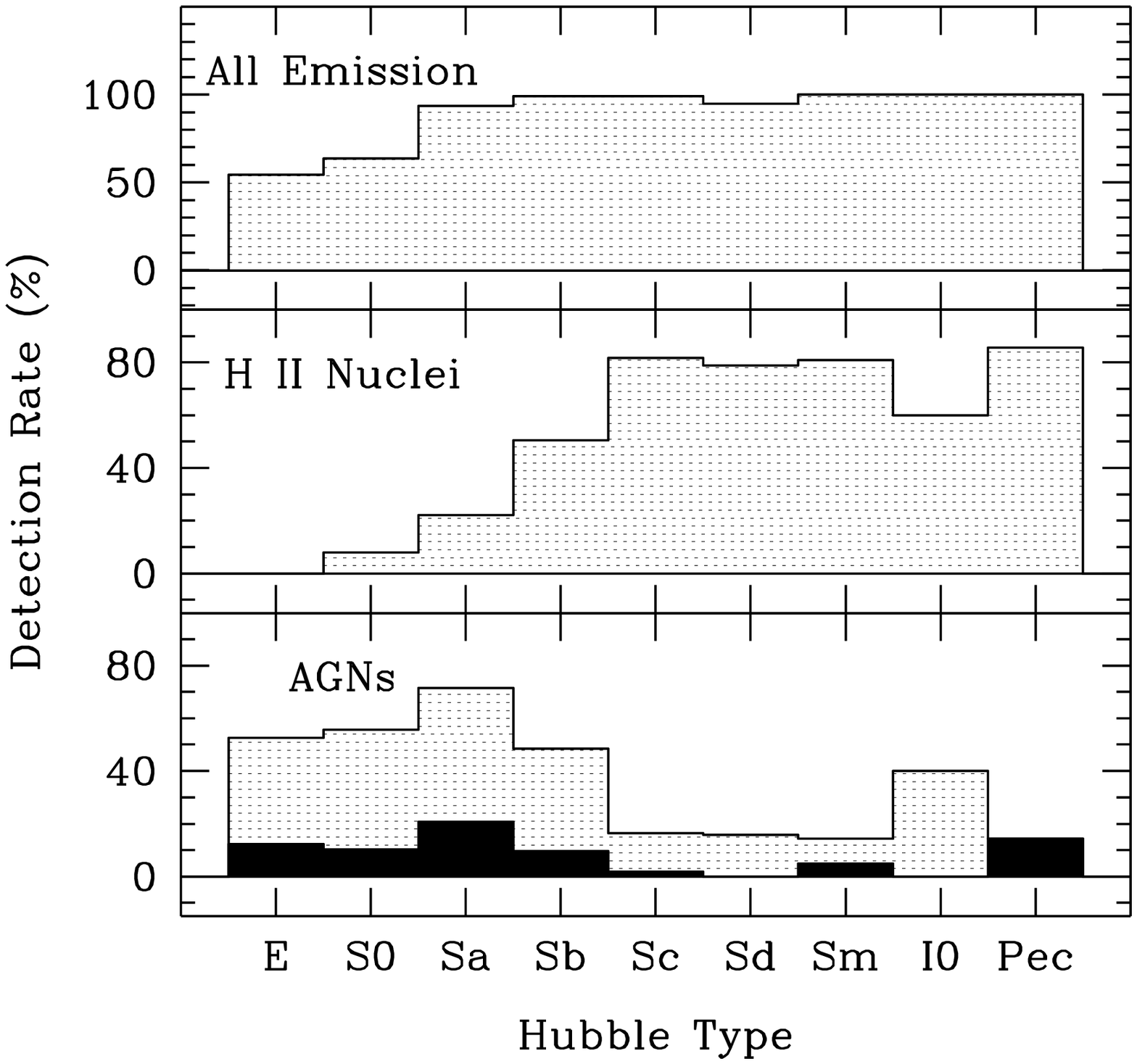,width=5.5cm,angle=0}
\psfig{figure=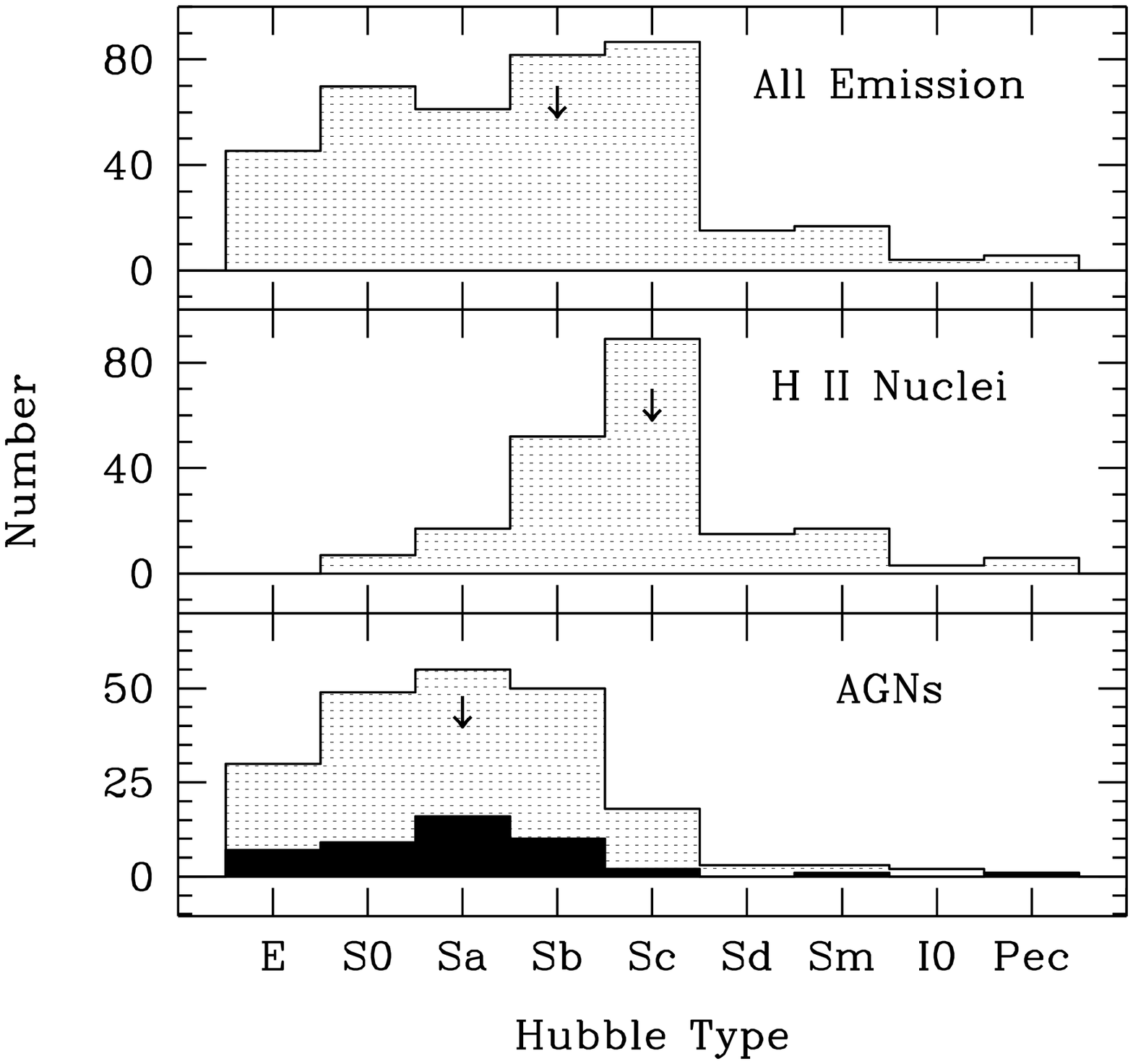,width=5.5cm,angle=0}
}
\caption{{\bf Left}
Detection rate as a function of Hubble type of all emission-line
nuclei, HII  nuclei, and AGNs
{\bf Right}
Distribution of morphological types for all emission-line nuclei,
HII nuclei, and AGNs; the downward-pointing arrow marks the
median of the distribution
(from Ho et al (1997)}
\label{fig3}
\end{figure}

In summary, the following points may explain
that there is no significant correlation between bars and nuclear
activity:
\begin{itemize}

\item to have an AGN a bar is not sufficient, there must already 
exist a massive black hole (which favors early-type objects)

\item the fueling requires a large central mass concentration

\item different time-scales  between the bar growth  and AGN 
activity cycles

\item there exist other fueling mechanisms like dense nuclear star clusters
(accounting for a time delay between the bar and fueling)
\end{itemize}

\section{Galaxy interactions and mergers}

Again, a strong correlation appears between AGN
and the presence of tidal interactions only for very luminous cases
(and accretion rates a few M$_\odot$/yr).
QSOs appear to interact with companions more than usual
(Hutchings \& Morris 1995, Bahcall et al  1997).
Seyferts can be fueled by 0.01 M$_\odot$/yr 
(like NGC 1068 for instance), and do not require external
interactions.

The role of the interactions is to trigger gravitational
instabilities in the galaxy disks, and the
 final fueling mechanism is a bar perturbation. Again
the fueling is favored for early-type objects.
The stability of the galaxies against perturbations
depends essentially on the bulge/disk ratio (e.g. Mihos et al 1996),
and the late-type objects are more susceptible to experience
violent flows leading to starbursts. The fueling of AGN
awaits the building of large mass concentration, the presence of
ILRs, and an efficient fueling of the very center.

\subsection{Frequency of companions}

Early studies (Dahari 1984, 1985)
found a correlation between AGN and the
presence of companions:
15\% of Seyferts have companions, compared
to 3\% of normal galaxies. While galaxies with companions
present no more H$\alpha$ or radio power, they are
more infra-red bright (McKenty et al 1989).
Keel et al (1985) remark the presence of 
 25\% of Seyfert in the close pairs of Arp Atlas, while
the frequency of Seyfert is only  5\% in a control sample.
 But the selection of the brightest objects in Arp Atlas
might bias the conclusions. Other studies 
(Bushouse 1986, DeRobertis et al 1996) find a deficiency of 
Seyferts  in interacting galaxies.
A paradoxical result is the large fraction of Seyfert in the low
surface brightness
(LSB) galaxies (Sprayberry et al 1995), since they are unevolved objects
generally in isolated environment. But the involved samples are small.

In a recent study, Schmitt (2001) does not find any
statistical difference in companion frequency for systems of the various
activity types. It has been claimed that
Seyfert 2 have a larger number of companions than Seyfert 1
(Laurikanen \& Salo 1995, Dultzin-Hacyan et al 1999),
but this is controversial (Schmitt et al 2001).
There could be an indirect link, some Seyfert 2 have an UV excess due to a
starburst (Cid Fernandes et al 1998), and companions enhance
the star formation.
Narrow-line Seyf1 also do not reveal more companions
(Krongold et al 2001).
The percentage of galaxies with companions  may sometimes 
appear higher for LINERs, or absorption-line galaxies
with respect to Seyfert and H II galaxies. But when only galaxies of similar 
morphological types are considered, this difference vanishes.
This suggests that the difference is due to a morphology-density effect
(Schmitt 2001).

\subsection{Compact groups}

Shimada et al (2000) find no statistical difference between
compact group galaxies and field galaxies, as far as AGN is concerned,
and conclude  that
 interactions do not trigger AGN activity nor starbursts.
 Wu et al (1998) draw the opposite conclusion, but for
very luminous infrared galaxies.
From their recent survey of 193 galaxies in 49 compact groups, 
Coziol et al (2000) confirm that AGN are located in the most luminous galaxies;
AGN prefer the early-type galaxies and starbursts the late-types.
AGN is the most frequent (41\%) type of activity; when
counting also the Starburst/AGN, the frequency rises 
to about 70\%.

\subsection{Radio Galaxies}

Here also, interactions are obvious in the more powerful objects.
FR-I have low-power, radio jets declining with radius; generally
hosted by elliptical galaxies, they have rare
interactions or tidal tails (less than 10\%, Smith \& Heckman 1989).
FR-II are high power, double-lobe sources like Cygnus-A, strong in their
lobe extremities: they show a high percentage of interactions
from 32 to 100\%  (Hutchings 1987, Yates et al 1989).
Most of them have two companions, and many star-forming regions.

Radio-loud QSOs have 4-5 times more neighbours, while
radio-quiet 2 times more than a control sample.
QSO are morphologically perturbed in 35-55\% of the cases,
radio-loud  QSO in 70-80\%
(Disney et al 1995,  Kirhakos et al 1999).

\subsection{Starburst - AGN Connection}

The role of gravitational interactions in luminous and ultra-luminous
infrared galaxies
have been established (Sanders \& Mirabel  1996).
In lower luminosity galaxies, only weak correlations are found
(Krongold et al 2001).
It appears that AGN and starbursts are two different phases of
the same evolutionary process.
In general Seyf2 have the most frequent starbursts
 (Levenson et al 2001).

Active galaxies have circumnuclear gas, it is correlated with star 
formation, and interactions (Storchi-Bergmann et al 2001)
and with « inner Hubble » types from Malkan et al (1998).
 High spatial resolution in the center of galaxies have 
emphasized the differences between Seyf1 and Seyf2, 
that do not have the same hosts statistically. Seyferts 2
possess more circumnuclear dust, and their ``inner'' morphological
types are shifted towards later type than the Seyferts 1.
Some of the dust obscuration characteristic of Seyf2 could come from 
this galactic circumnuclear dust more than from an accreting torus
(Malkan et al. 1998).

\begin{figure}[t]
\centerline{
\psfig{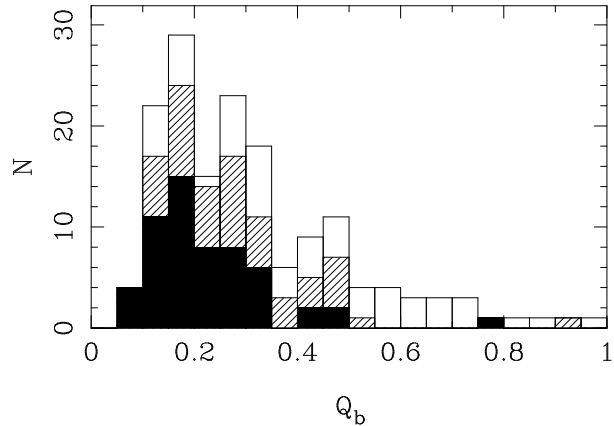}
}
\caption{Histogram of gravitational bar strengths in the
OSU sample. The shading indicates the de Vaucouleurs bar family:
SA (black), SAB (hatched), and SB (white);
from Block et al (2002)}
\label{fig4}
\end{figure}

\section{Orientation of accretion disks}

To better understand the fueling mechanisms, it might be
interesting to investigate the orientation of accretion disks
observationally, with respect to the orientation of the galaxies
themselves. Since the accretion disks are too small to
be observed directly, a common tracer are
radio jets, that are perpendicular to the accretion disks, 
and give their orientation.  
Their orientation has been compared
with the rotation axis of the spiral galaxy host, but 
random orientations have been found (Clarke et al, 1998, 
Nagar \& Wilson 1999, Kinney et al 2000).

Radio galaxies (in general of elliptical type) have
been found to have jets perpendicular to their dust lanes 
(Kotanyi \& Ekers 1979, van Dokkum \& Franx 1995, 
de Koff et al 2000, for FRI only).
But this is put into question in a recent study of 20 radio
galaxies (Schmitt et al 2002): jets appear to have rather
random orientation between 55$^\circ$ to 77$^\circ$ from the 
galaxy disks, while they appear to avoid an orientation too close
to the disks. They are then ``only roughly'' perpendicular to the 
disks, but do not align strictly with the disk axis.
This might not indicate a random orientation of the fueling through
mergers and gas accretion, but this misalignement could be due to
warping instabilities of the accretion disks, 
or warping of galaxy disks.

\section{Conclusions} 

There is no correlation between AGN activity and presence of bars
or interactions, except for the high luminosity objects.
Many of the correlations claimed come from morphological type
mismatch between AGN and control sample.
AGN are predominantly found in early-type galaxies, while
starbursts are found in late-types.
But precisely this has to be taken into account! Since the morphology 
of galaxies change during their evolution.

A large set of observations compared to numerical simulations
have shown that bars are only transient in galaxy disks, when
gas is present. Bars produce radial inflows of gas able to destroy
them, but they can re-form through gas accretion (Bournaud \& Combes 2002).
When there exists a central mass concentration (the galaxy looks more
« early »), it is more easy to fuel the nucleus.
When gas is accreted all over the disk, starbursts are triggered, 
and the galaxy looks more late-type, but is not yet an AGN.
A galaxy is in continuous evolution, and accretes mass all along
its life. The evolution is governed by
self-regulated processes, through bars and interactions. 

The bulge-to-disk ratio and the gas fraction can evolve
and therefore the morphological type  somewhat 
oscillates from early to late-type, while secularly evolving
to early-types.
Any galaxy will be barred, active, 
 and spend some time as an early-type or late-type; the frequency
of bars in galaxies can be interpreted within this evolution
(cf fig 4, Block et al 2002).
It is when the gas has accumulated towards the center, when the type
is early (although the ``inner'' type is more late) that nuclear activity
is triggered.
The morphological-type/AGN correlation is a consequence of this
scenario.

\vspace{-3mm}

\small{

}
\end{document}